# Digitally reconfigurable complex 2D Dual lattice structure by optical phase engineering

Manish Kumar* and Joby Joseph

*Photonics Research Laboratory, Department of Physics, Indian Institute of Technology Delhi, New Delhi, India - 110016*
*Corresponding author: manishk.iitd@gmail.com*

We present a method to combine two periodic lattice wave-fields to generate a complex dual lattice wave-field which could be employed for microfabrication of corresponding 2D dual lattice structures. Since the addition of two periodic lattice wave-fields is coherent in nature the resultant dual lattice structure is highly dependent on the relative phase difference between constituent wave-fields. We show that it is possible to have control over the dual lattice pattern by precisely controlling this relative phase difference. This control is enabled by making use of digitally addressable phase only spatial light modulator (SLM). We provide the computational method for calculation of corresponding phase mask to be displayed on the SLM and also verify the results experimentally by employing a simple 4f Fourier filter based geometry. The method is completely scalable and reconfigurable in terms of the choice of periodic lattice wave-fields and has the potential to form gradient phase masks which could be useful for fabrication of graded index optical components.

## 1. Introduction

Photonic Crystals (PhCs) are periodic dielectric (or metallic) structures capable of modifying the flow of light [1]. Depending on the periodicity along one, two or three orthogonal directions, they are categorized as 1D, 2D or 3D PhCs. In present work our focus is on the case of 2D PhCs. 2D PhCs are useful in all optical circuit waveguides [2], micro lasers [3], and in some special effects like superprism [4], supercollimation [5], negative refraction [6], slow light [7] etc.

PhCs may be fabricated by a number of methods, such as direct laser writing [8], self-assembly method [9], e-beam lithography [10], and multiple beam interference lithography based methods [11]. The interference lithography based method is of particular interest as this is a single shot large area patterning method. The multiple beams could be generated either by using multiple beam splitters [12] or by using a phase only diffractive optical element (DOE) [13]. The phase DOE based method is interesting because of its experimental simplicity when compared to the use of multiple beam splitters. Moreover with the implementation of DOE, by making use of a digitally addressable phase only spatial light modulator (SLM), it is possible to control the relative phase difference between the interfering beams. Such control over the phase difference has enabled the fabrication of complex structures such as chiral photonic crystals [14] and even has enabled embedding of defect sites in such complex structures [15]. Recently, Lutkenhaus et.al. [16] showed by using diffraction theory that it is possible to produce a phase mask which, when displayed on a SLM, can control the relative phase shift between diffracted beams. They did it by mapping the 2D space with hexagons where each hexagon is made of six equilateral triangles. The control over grey level of these triangles produce the required phase mask where a gradual change in grey levels over the triangles shows up as a change in relative phase difference between diffracted beams. This control over relative phase difference was verified by the visible change in dual lattice structure formed by interference of selected four rotationally non-symmetric beams. They further pointed that it is possible to extend the ideas to generate a gradient phase mask which could be helpful in realizing graded index photonic crystal structure.

In this work, we propose an alternative approach to generate the phase mask required to obtain such non-symmetrical beams to form a dual-lattice structure. Our method is very much based on the phase engineering approach presented in some recent works [15, 17]. In our approach we visualize the dual-lattice structure as combination of a grating structure with hexagonal lattice structure and produce it by coherent superposition of a 2-beam wave-field and a 3-beam wave-field in such a way that one of the plane waves are common in both the constituent wave-fields. This method is advantageous in terms of more efficient use of available laser power and is more flexible in terms of phase assignment. The calculation involved, to get the phase mask for changing the relative phase difference, is very straight forward. We show that our approach, even while making use of the same 4 non-symmetrical beam interference geometry as in ref. 16, gives rise to more variety of phase dependent dual-lattice patterns. Moreover, our method being very general is easily extendable to other complex geometries of such dual-lattice, or even more than two lattice, structures.

## 2. Experimental setup

We use a diode laser (Toptica BlueMode, Germany) emitting at 405nm wavelength, a 20x microscope objective and lenses with focal lengths of 135 mm (for collimation), f1 = 500 mm and f2 = 135 mm (for 4f Fourier filter setup), and a phase-only SLM (Holoeye-LETO, Germany), which is a reflective LCOS microdisplay with 1920x1080 pixel resolution and 6.4µm pixel pitch. Fig. 1 shows the schematic of experimental setup. The numerically computed phase masks were displayed on the SLM so that the incident collimated laser beam, after getting reflected from SLM, gets modulated as per the displayed phase masks and generates multiple diffracted beams. A Fourier filter was used in the Fourier plane to select only the desired four non-symmetric beams. The second lens controls the demagnification factor in this arrangement. A CMOS camera (UI-1488LE-M IDS, Germany) was used to record the interference pattern in the imaging plane.

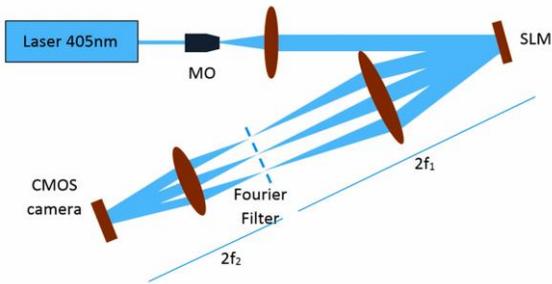

Fig. 1: (Color online) Schematic diagram for the experimental setup. MO: microscope objective, SLM: spatial light modulator (phase only).

### 3. Design of phase mask for dual-lattice wave-field

Our purpose here is to describe the method of designing a phase mask for producing four non-symmetrical beam interference.

#### A. Dual lattice wave-field as the sum of two periodic wave-fields

When all the interfering beams make same angle with normal to the interference plane, the resultant wave-field is essentially non-diffracting in nature. This is further verified by the irradiance profile at Fourier transform plane showing that all the required wave-vectors are spread along a cone. It is known that, coherent superposition of two or more non-diffracting wave-fields is non-diffracting in nature [15, 17]. This statement implies that it could be possible to split a complex non-diffracting wave-field into coherent sum of simpler non-diffracting wave-fields. In present work, we propose that one could split a dual lattice wave-field into superposition of two periodic lattice wave-fields. In particular we noticed that the dual lattice wave-field, under our consideration [16], could be seen as coherent sum of two wave-fields namely a 2-beam interference wave-field and a 3-beam interference wave-field. We represent it by the following Eq. (1). Eq. (2) shows the construction of individual n-beam interference wave-fields where $E_i$, $k_i$, and $r$ represent complex field amplitudes, wave vectors, and position vector for the interfering plane waves respectively.

$$E_{dual-lattice} = E_{2-beam} + E_{3-beam} \quad ....(1)$$

$$E_{n-beam} = \sum_{i=1}^{n} E_i e^{i k_i \cdot r} \quad ....(2)$$

One may arrive to this conclusion even by considering the picture in the Fourier domain. Fig. 2(a)-(c) does the schematic illustration of how sum of two constituent lattice wave-fields with one common plane wave component effectively results into 3+2-1=4 non-symmetric beam interference arrangement to give rise to dual lattice wave-field. In Fig. 2(d)-(f) we show the corresponding irradiance profile due to interference of the beam geometry corresponding to Fig. 2(a)-(c) respectively. The numerical simulation results for corresponding interference profiles further confirmed that dual lattice wave-field is indeed made of superposition of two periodic wave-fields. These numerical simulation results for multiple beam interference were obtained by making use of following equation

$$I(r) = \sum_{i=1}^{n} |E_i|^2 + \sum_{i=1}^{n} \sum_{\substack{j=1 \\ j \neq i}}^{n} E_i E_j^* \cdot \exp\left[i(k_i - k_j)r + i\psi_{ij}\right] ....(3)$$

where $\psi_{ij}$ represents the difference in initial phase offsets of the interfering beams. Index i=1 to n corresponds to the n side beams and $k_i$ is the wave-vector corresponding to ith side beam which is given by

$$k_i = k * [\cos(q_i \pi) * \sin\theta_i, \sin(q_i \pi) * \sin\theta_i, \cos\theta_i] \quad ....(4)$$

where $k = 2\pi/\lambda$ ($\lambda$ being the wavelength of laser in air), $q_i = 2(i-1)/n$ and $\theta_i$ is the angle which ith side beam makes with normal to the interference plane. The magnitude of each of the complex field amplitudes is taken to be unity.

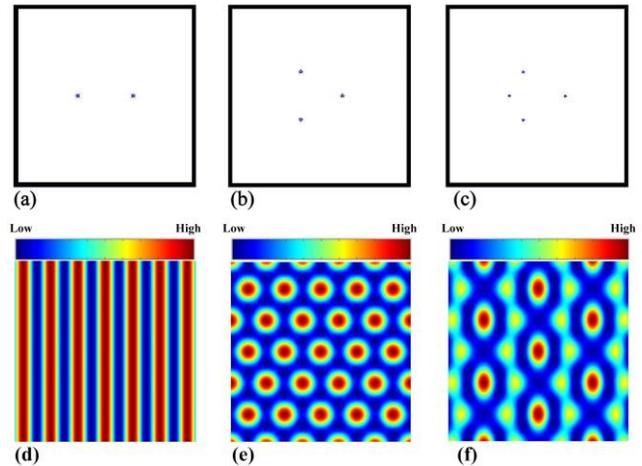

Fig. 2: (Color online) Multiple beam interference patterns. Fourier space picture representing: (a) 2-beam arrangement, (b) 3-beam arrangement, and (c) sum of both 2-beam and 3-beam arrangements where one beam is in common. (d)-(f) numerically calculated interference pattern due to the beam arrangements shown in (a)-(c) respectively.

#### B. Synthesis of the phase mask

Having confirmed that dual lattice wave-field under consideration is made of superposition of grating and hexagonal wave-field, it is very simple to synthesize the corresponding phase mask. We noted that a class of non-diffracting beam geometry can efficiently be encoded just by using a synthetic phase hologram which is nothing but the phase only component of the corresponding complex wave-field [18]. This method has effectively been used for encoding a variety of non-diffracting fields [13, 15, 17] and is applicable in our present work as well. The extraction of phase mask for dual lattice wave-filed is represented in Fig. 3 where (a)-(c) help visualize the process of obtaining complex wave-field $E_{dual\text{-}lattice}$ by making use of Eq. (1) where $E_{2\text{-}beam}$ and $E_{3\text{-}beam}$ were obtained by making use of superposition of plane waves as described by Eq. (2).

Inset images in Fig. 3(a)-(c) represent corresponding Fourier transform intensity patterns obtained by making use of FFT algorithm in MATLAB. In Fig. 3(d) is the extracted synthetic phase which is the required phase mask for encoding the dual lattice wave-field by making use of a SLM in the experimental setup. Fig. 3(e) shows numerically computed FT of this phase mask. Dual-lattice wave-field is formed by interference of the beams corresponding to spots numbered as 1, 3, 4 and 5 which are selected by making use of a Fourier filter in the Fourier plane. One advantage of our method is very clear from here that, our phase mask being synthesized by the method introduced by Arrizon et.al. [18], is very efficient and directs most of power in desired four spots against the phase mask obtained in ref. 16, where even the spot numbers 2 and 6 are equally intense as other spots. Having maximum power in the desired interfering beams is of practical interest when recording the interference pattern in a photosensitive medium.

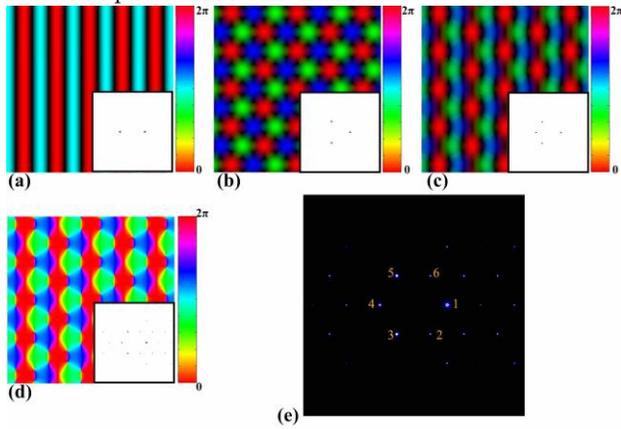

Fig. 3: (Color online) Complex wave-fields visualization for (a) 2-beam interference, (b) 3-beam interference and (c) coherent sum of 2-beam and 3-beam interference. The color represents the phase distribution of complex field according to the colorbars associated with each figure while brightness varies linearly with amplitude. (d) Extracted phase profile from the complex field shown in (c). Insets in (a)-(d) show numerically obtained Fourier transform intensity profile of corresponding complex/phase profiles. (e) Enlarged view of Fourier transform of phase profile of (d). Spots are numbered from 1 to 6 where spot number 1, 3, 4 and 5 are responsible for forming dual-lattice wave-field.

### C. Tunable dual lattice wave-field

In last section we described how we obtained the phase mask for dual-lattice wave-field in a very straight forward manner. Next we want to have a method to be able to control the relative phase difference between the interfering beams to have a visible control over the dual-lattice pattern hence making them digitally tunable. Lutkenhaus et.al. [16] achieved this by encoding the phase mask to have control over the relative phase difference between two beam sets where set-1 consisted of beam numbers 1, 3 and 5 while set-2 consisted of beam number 2, 4 and 6. In our approach, since we superpose two periodic lattice wave-fields, we can easily modify the phase difference between these two constituent periodic lattice wave-fields as described by Eq. (5)

$$\mathbf{E}_{dual-lattice} = \mathbf{E}_{2-beam} + \mathbf{E}_{3-beam} \times e^{i\varphi} \quad \ldots(5)$$

where factor $\varphi$ is the relative phase difference between constituent wave-fields and it controls the change in the pattern of the dual-lattice wave-field. In order to see the effect of change of $\varphi$ we synthesized the phase mask for different values of $\varphi$ ranging from 0 to $2\pi$ and also computed the corresponding dual-lattice interference patterns numerically. The results are compiled in Fig. 4 along with corresponding experimentally obtained Fourier transforms and interference patterns as well.

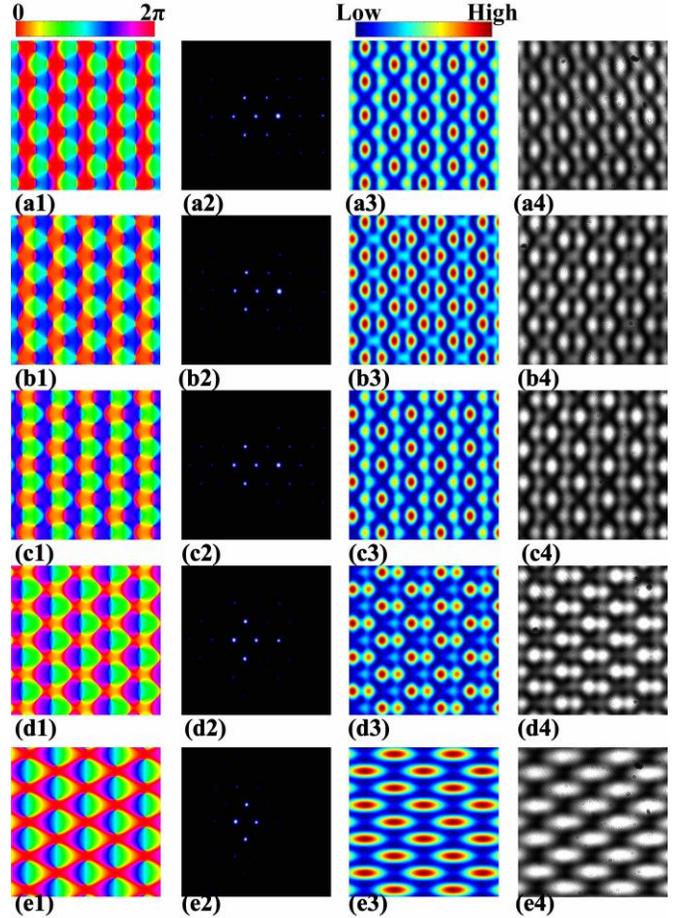

Fig. 4: (Color online) (Left to right) Dual-lattice wave-field phase profile (1st column), experimentally obtained Fourier transform intensity profile (2nd column), numerically simulated interference intensity profile (3rd column) and experimentally recorded interference intensity profile (4th column). Row numbered (a) to (e) represent the results for different phase differences between coherent sum of 2-beam and 3-beam wave-fields: 0, π/4, π/2, 3π/4 and π respectively.

We have skipped displaying the results corresponding to $\varphi$ in the range of π to 2π as it produced same results as the range π to 0. It is clear that in our method $\varphi$ = 0 to π/4 (or alternatively π/2 to π/4) range covers the whole pattern range as obtained in ref. 16. In addition, range $\varphi$ = π/2 to π produces a gradual shift of dual lattice pattern from a fenced hexagonal structure to a dumb-bell shape hexagonal lattice to hexagonal lattice with elliptical basis. Here we wish to point out that this particular range of phase shift, i.e. $\varphi$ = π/2 to π, offers greater variety of phase dependent pattern change. While doing micro-fabrication in a photoresist (or other light sensitive medium) this would enable one to control the overall fill factor of written pattern by controlling the phase φ in this phase range.

In Fig. 4 we also display the corresponding Fourier transform intensity profiles as captured, on the Fourier plane of first lens of 4f setup, by a DSLR camera (Nikon-D5100) and it is clearly seen that one of the spots, which corresponds to the common wave-vector among both the constituent periodic wave-fields i.e. spot number 1, changes in intensity as the phase is changed from 0 to π. At φ = 0 this spot is twice as intense as other contributing spots and at φ = π, its intensity goes to zero confirming the exact phase response as expected. There is an additional spot at center due to reflection of some laser power from SLM-air interface which is easily removed during the process of Fourier filtering. Experimentally obtained intensity profiles in Fig. 4 (a4)-(e4) were captured by CMOS camera with experimental setup having f1 = f2 = 500mm. Actual numerical calculations for phase mask synthesis was done for $\theta_i$ = 0.4°, over 1920x1080 pixels of SLM with 6.4μm pixel size.

### 4. Micro-fabrication in photoresist medium

Next we did micro-fabrication of one of the intensity profiles Fig. 4(a3) in the photoresist medium to show the applicability of our method in such fabrication processes. We used a positive photoresist (AZ1518) in our experimental setup with 405nm diode laser (Toptica BlueMode) and f1 = 500mm, f2 = 135mm Fourier transforming lenses. The photoresist was spin coated (using Spin NXG-P1, Apex Instruments, India) on a glass slide at 4000 RPM for 60 seconds to obtain approximately 2μm thick layer. The photoresist film was prebaked on a hot-plate (EchoTherm HP50, Torrey Pines Scientific) for 50 seconds at 100°C. Pre-baked sample was exposed to the interference pattern for 9 seconds. The exposed sample was then developed in the developer solution (AZ726 MIF) for 60 seconds. Developed sample showed nicely written uniform pattern as shown in Fig. 5. Fig. 5(a) is the image obtained at 40x magnification by a microscope and in Fig. 5(b) we have a scanning electron microscope (SEM) image by using a benchtop SEM (JEOL NeoScope JCM6000). Theoretical calculation for demagnification factor of f1/f2 = 500/135 with $\theta_i$ = 0.4° (for the DOE loaded on SLM) predicts the lattice constant of 18.1μm which is in good agreement with the experimentally obtained value shown in Fig. 5(b).

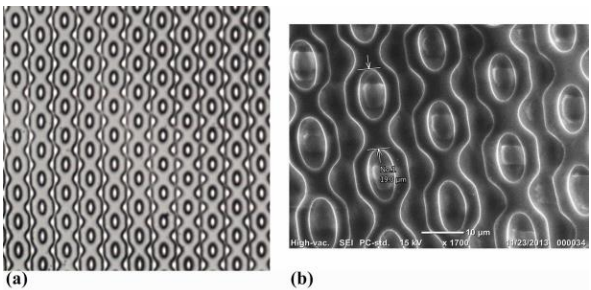

Fig. 5: Experimentally recorded dual-lattice pattern in a positive photoresist. (a) An image obtained by optical microscope and (b) SEM profile with corresponding length scale bar.

### 5. Design of a Gradient phase mask and discussion

It is clear from section 3 that dual-lattice interference pattern is highly dependent on phase difference φ and this phase difference could precisely be controlled by synthesizing the phase mask with a given phase difference. This change in pattern gets reflected into a change in fill-factor of the recorded pattern in photoresist. Thus it is possible to control the fill factor in a material by controlling the relative phase difference φ of the corresponding phase mask. A spatially varying fill factor could possibly be obtained by using a spatially varying phase mask. Such a phase mask could be synthesised by replacing a uniform φ with a space dependent φ(x,y) in Eq. (5). A spatially varying fill-factor structure could be applicable as a graded lattice/index material [19] leading to many applications. Fig. 6 schematically summarises the synthesis of one such graded phase mask for the proposed dual-lattice structure where φ is varied in linearly fashion along the radius from value π/2 (at origin) to π (at corner).

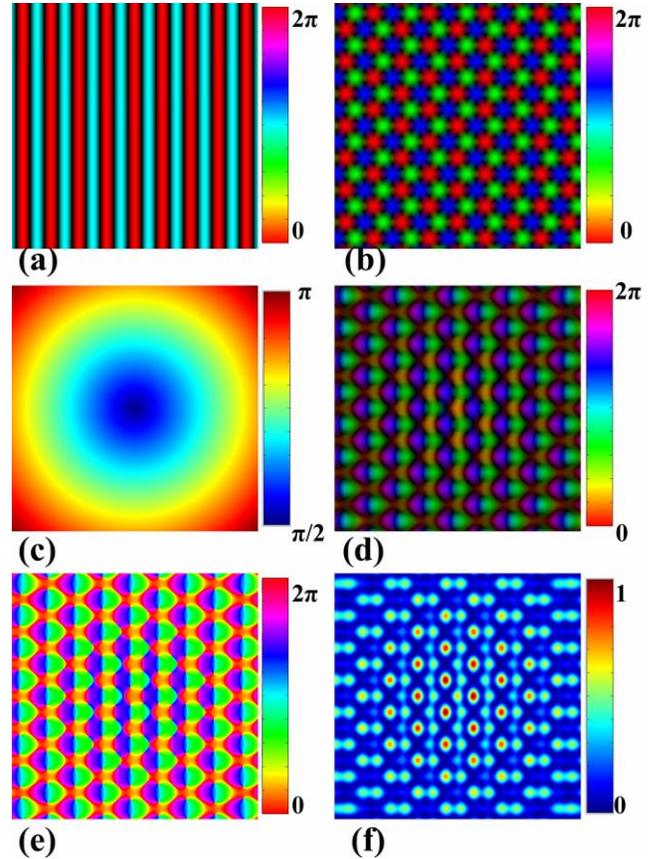

Fig. 6: (Color online) Synthesis of a gradient phase mask. Complex wave fields due to (a) 2-beam and (b) 3-beam interference. (c) Spatially varying phase profile which varies linearly in the radial direction from value π/2 to π. (d) resultant complex wave-field due to superposition of wave-fields shown in (a) and (b) where initial phase offset between fields is given by (c). (e) Extracted phase profile from the resultant complex field (d). (f) Intensity profile obtained from complex wave-field in (d).

It is worth noticing that our proposed method of forming dual-lattice wave-field is completely reconfigurable and general. It is not restricted to the combination of a 2-beam and a 3-beam wave-field alone as it could easily be extended to any set of non-diffracting wave-fields.

### 6. Summary

In summary, we have given a direct and generalized method for synthesising phase mask for forming a dual lattice structure. The synthesised phase mask splits an incident plane wave into multiple non-symmetrical plane-wave geometry with intended initial phase offsets. These beams are then made to interfere by making use of a 4f Fourier filtering geometry to produce a dual-lattice wave-field which was recorded into a photoresist medium. As an example we have shown the formation of dual lattice structure by interference of four non-symmetrical plane waves. The periodicity of structure could be controlled by changing the focal length ratio of two lenses used in 4f geometry. The presented method could be useful for fabricating photonic lattice based gradient index devices and is also easily extendable to other more complex beam geometries.

Manish Kumar wishes to acknowledge the financial assistance from Council of Scientific and Industrial Research (CSIR), New Delhi, India.